\documentclass[sigconf]{acmart}

\usepackage[utf8]{inputenc}


\usepackage{xspace}
\usepackage{multirow}
\usepackage{booktabs}

\usepackage{xargs}

\usepackage{listings}
\usepackage{./lstlinebgrd}

\newcommand{\Web}{Web}
\newcommand{\EG}{e.g.}
\newcommand{\IE}{i.e.}
\newcommand{\JS}{JavaScript}
\newcommand{\EtAl}{et al.}
\newcommand{\EL}{EasyList}
\newcommand{\ROCAUC}{ROC-AUC}
\newcommand{\XGBOOST}{XGBoost}
\newcommand{\ML}{ML}
\newcommand{\PG}{PageGraph}
\newcommand{\BB}{Brave Browser}
\definecolor{dkgreen}{rgb}{0,0.6,0}
\definecolor{gray}{rgb}{0.5,0.5,0.5}
\definecolor{mauve}{rgb}{0.58,0,0.82}
\definecolor{editorGray}{rgb}{0.95, 0.95, 0.95}
\definecolor{editorOcher}{rgb}{1, 0.5, 0}
\definecolor{editorGreen}{rgb}{0,0.6,0}
\definecolor{darkgreen}{RGB}{0,90,90}
\definecolor{lightgray}{rgb}{0.95, 0.95, 0.95}
\definecolor{darkgray}{rgb}{0.4, 0.4, 0.4}
\definecolor{orange}{rgb}{1,0.45,0.13}		
\definecolor{olive}{rgb}{0.17,0.59,0.20}
\definecolor{brown}{rgb}{0.69,0.31,0.31}
\definecolor{purple}{rgb}{0.38,0.18,0.81}
\definecolor{lightblue}{rgb}{0.1,0.57,0.7}
\definecolor{lightred}{rgb}{1,0.4,0.5}
\definecolor{dkgreen}{rgb}{0,0.6,0}
\definecolor{gray}{rgb}{0.5,0.5,0.5}
\definecolor{mauve}{rgb}{0.58,0,0.82}
\definecolor{editorGray}{rgb}{0.95, 0.95, 0.95}
\definecolor{editorOcher}{rgb}{1, 0.5, 0}
\definecolor{editorGreen}{rgb}{0,0.6,0}

\definecolor{MyMaroon}{HTML}{D50000}
\definecolor{MyYellow}{HTML}{FFD600}
\definecolor{MyGreen}{HTML}{00C853}

\definecolor{MyTeal}{HTML}{009688}
\definecolor{MyOrange}{HTML}{FF5722}
\definecolor{MyGray}{HTML}{E0E0E0}
\definecolor{MyIndigo}{HTML}{3F51B5}
\definecolor{MyPurple}{HTML}{9C27B0}
\definecolor{MyBlack}{HTML}{000000}
\definecolor{MyBlue}{HTML}{3F51B5}

\newcommand{\SubPoint}[1]{\textbf{#1}. }

\newcommand{\SourceLink}{\textit{Link to source code omitted for anonymization.}}
\newcommand{\PrivacyContributionsLinks}{\textit{Links and references omitted for anonymization.}}
\newcommand{\NumDataPoints}{2,662}

\newcommand{\NumTrainingSamples}{1,966}

\newcommand{\NumPositiveLabels}{1,011}

\newcommand{\NumNegativeLabels}{955}

\newcommand{\NumFeatures}{433}

\newcommand{\NumOuterFolds}{10}

\newcommand{\NumInnerFolds}{3}

\newcommand{\NumFeatureImpFolds}{5}

\newcommand{\NumOptimisationSteps}{10}

\newcommand{\NullValueThreshold}{0.73}

\newcommand{\CorrelationThreshold}{0.85}

\newcommand{\ClassifierMeanAUC}{0.88}
\newcommand{\ClassifierAUC}{\ClassifierMeanAUC}
\newcommand{\ClassifierStdAUC}{0.03}

\newcommand{\NumImportantFeatures}{40}

\newcommand{\ConvergedAfterNumSamples}{983}
\newcommand{\ConvergedAfterPercentageSamples}{50}

\newcommand{\MeanLossAUCForIntervention}{0.004}
\newcommand{\MeanLossAUCForPage}{0.010}
\newcommand{\MeanLossAUCForExpert}{0.004}
\newcommand{\MeanLossAUCForAuto}{0.037}
\newcommand{\MeanLossAUCForAbsolute}{0.024}
\newcommand{\MeanLossAUCForRelative}{0.009}

\newcommand{\NumFilterListRules}{1,469}

\newcommand{\NumWebPages}{2,570}
\lstdefinestyle{code} {%
  backgroundcolor=\color{editorGray},
  basicstyle=\fontsize{7}{7}\ttfamily,
  frame=tb,
  captionpos=b,
  stepnumber=1,
  firstnumber=1,
  numberfirstline=true,	
}

\usepackage{tikz}
\usetikzlibrary{arrows.meta}
\usetikzlibrary{positioning}
\tikzset{%
  >={Latex[width=2mm,length=2mm]},
        base/.style = {rectangle, rounded corners, draw=black,
                           minimum width=1cm, minimum height=1cm,
                           align=center, font=\sffamily},
        parser/.style = {fill=red!10},
        dom/.style = {fill=blue!10},
        resource/.style = {fill=yellow!10},
        sect/.style = {rectangle, draw=black,
                   minimum width=8.2cm, minimum height=1cm,
                   align=center, font=\footnotesize\sffamily, fill=yellow!10},
        user/.style = {rectangle, draw=white,
                       align=center, font=\footnotesize\sffamily},
        result/.style = {rectangle, rounded corners, draw=black,
                           minimum width=1cm, minimum height=1cm,
                           align=center, font=\footnotesize\sffamily, fill=pink!20},
}

\begin{document}

\author{Michael Smith}
\affiliation{
    \institution{University of California, San Diego}
    \country{United States of America}
}
\email{mds009@eng.ucsd.edu}

\author{Peter Snyder}
\affiliation{
    \institution{Brave Software}
    \country{United States of America}
}
\email{pes@brave.com}

\author{Moritz Haller}
\affiliation{
    \institution{Brave Software}
    \country{United States of America}
}
\email{mhaller@brave.com}

\author{Benjamin Livshits}
\affiliation{
    \institution{Imperial College London}
    \country{United Kingdom}
}
\email{b.livshits@imperial.ac.uk}

\author{Deian Stefan}
\affiliation{
    \institution{University of California, San Diego}
    \country{United States of America}
}

\author{Hamed Haddadi}
\affiliation{
    \institution{Brave Software}
    \country{United States of America}
}
\email{hamed@brave.com}





\title{Blocked or Broken? Automatically Detecting When Privacy Interventions Break Websites}

\begin{abstract} 
A core problem in the development and maintenance of crowdsourced filter lists 
is that their maintainers cannot confidently predict whether (and where)
a new filter list rule will break websites. 
This is a result of enormity of the \Web{}, which prevents filter list authors from broadly understanding the impact of a new blocking rule before they ship it to millions of users. The inability of filter list authors to evaluate the \Web{} compatibility impact of a new rule
\emph{before shipping it} severely reduces the benefits of filter-list-based content blocking: filter lists are both overly-conservative (\IE{} rules are tailored narrowly to reduce the risk of breaking things) and error-prone (\IE{} blocking tools still break large numbers of sites).
To scale to the size and scope of the Web, filter list authors need
an automated system to detect when a new filter rule breaks websites,
before that breakage has a chance to make it to end users.
\\
\hphantom{-----}In this work, we design and implement the first automated system for predicting
when a filter list rule breaks a website. We
build a classifier, trained on a dataset generated by a combination of compatibility data
from the EasyList project and novel browser instrumentation, and
find it is accurate to practical levels (AUC \ClassifierMeanAUC{}).
Our open source system requires no human interaction when assessing the compatibility risk of a proposed privacy intervention. We also present the 40 page behaviors that most predict breakage in
observed websites.%
\end{abstract} 

\maketitle{}

\section{Introduction}

A large and growing body of research has shown that filter list based
content blocking significantly improves Web privacy\cite{merzdovnik2017block, gervais2017quantifying},
security\cite{li2012knowing, zarras2014dark} and performance\cite{garimella2017ad, pujol2015annoyed}.
The popularity of cosmetic-only rules in popular filter lists suggests
that filter lists significantly improve the user-perceived aesthetics of
Web browsing. And the (large and growing) popularity of extensions and browsers
that include filter-list based content blocking suggests that
filter lists are important to large percentages of Web users.

While the benefits of filter lists are well studied and understood, 
systematizing and automating the creation of filter lists remains an open challenge.
This is largely
because research is very good at measuring the \emph{benefits} of blocking
network requests (\EG{} numbers of trackers blocked, data saved,
CPU cycles reduced), but comparatively poor at measuring the \emph{costs} of
blocking requests (\EG{} number of websites broken or user-desirable features
impacted).

In effect, Web researchers mainly count one side of the
ledger, and as a result, filter list curation in practice remains a nearly
completely manual process, consisting of activists and community members making
best effort predictions of the Web-scale impact of filter list rules.
The result is that filter lists are both too conservative (\IE{}
there are things that filter list authors would like to block, but don't to avoid breaking sites) and too liberal (\IE{} content blocking
tools still break lots of websites).

Additional manual effort (\EG{}
more filter list maintainers) will not fundamentally improve the situation
either. Because of the size and constant change of the Web, any efforts to
\emph{manually} evaluate the Web-wide impact of a filter list rule by filter
list authors is going to be incomplete (and dramatically so). As a result,
users of filter list tools---be they extensions or browsers---end
up being both the consumers and testers of new filter list rules. This means
broken sites for users, and in some case giving up on the privacy,
security and performance wins of content blocking tools.

\subsection{Problem Difficulty}

We need an automated way to predict the web-compat impact of
a filter list rule, so that rules can be tested, tailored and optimized
before being shipped to users. Alas, this is difficult problem for
several reasons. 


First, determining if a page is broken is difficult because
``brokenness'' can present itself in different ways, many
of which only reveal themselves after a user interacts with a page or
attempts to trigger some interactivity. A page could be ``broken'' in an
obvious way (\EG{} the page is blank), in a very subtle way (\EG{} a
form on a deeply nested page does not submit correctly), and everything in
between. The wide range of ways a site can break makes automated detection
difficult too.

And second, automated detection is difficult because its difficult to build
up a large dataset of ``broken'' web sites. Both site authors and filter list
maintainers have strong incentives to fix broken sites as quickly as possible
(and generally, with as few people noticing as possible). This makes it
difficult for researchers to get a generalized understanding of the problem,
and so makes developing automated detection systems tricky.

\subsection{Contributions}

This work improves the state of filter list content blocking by 
designing a fully automated classifier that
accurately predicts whether a filter
list rule break a website, in the subjective evaluation of a browser user.
Our classifier requires no human action
to run, and runs in a heavily-modified desktop \Web{} browsers, and so can
scale far beyond what is possible with human assessments. Our classifier
takes as input i) a filter list rule, and ii) a website URL, and returns a
prediction of whether executing the given website \emph{with the given
filter list rule applied} will break the website.

We build our classifier in two novel steps. First, we use the commit
history of the EasyList project to build up a labeled dataset of filter list
rules that did (and did not) break websites, in the subjective evaluation
of the users and maintainers of EasyList authors and users.

And second,
we use a heavily modified version of a Chromium-based browser to analyze
the execution of the website. Significantly, our modified browser (built
from an existing system called PageGraph\footnote{\url{https://github.com/brave/brave-browser/wiki/PageGraph}}) records both what events occurred
during the website's execution (\EG{} which scripts were executed, which
DOM notes were inserted or modified, which event listeners were registered),
and which actors on the page were responsible for each event (\EG{}
which script fetched a given resource, or inserted a DOM element, or fetched
a dependent script). Our PageGraph-instrumented browser then allows us
to export the recording of each page execution as a XML-encoded graph.

We then combine these sources of data to generate a large dataset of
website executions, where we know (or are highly confident) that the site
was working or broken, in the subjective determination of a browser user.
More specifically, we generate \NumDataPoints{}, each with a label of
\texttt{\{broken, working\} X \{``with rule applied'', ``without rule applied''\}}.
We then extract \NumFeatures{} features from each data point, and train
a classifier that performs with AUC of \ClassifierAUC{}.

More specifically, this work offers the following contributions:

\begin{enumerate}
  \item The design of a multi-step \textbf{fully automated system} for accurately predicting
    whether a privacy intervention (\IE{} a filter list rule) would break a website, in the subjective
    evaluation of a browser user.
  \item A \textbf{public dataset} consisting of \NumFilterListRules{} unique real-world filter list rules, applied to \NumWebPages{} unique \Web{} pages that they affect, resulting in \NumDataPoints{} recordings of page behavior changes, each labeled with whether the rules yielded a broken or working version of the page.
  \item A detailed \textbf{discussion of which page behaviors predicted pages breaking}
    (and which page behaviors did not).
  \item The \textbf{open source implementation}\footnote{\SourceLink} of both our data collection
    pipeline and resulting classifier, implemented in a Chromium-based browser
    and scikit-learn.
\end{enumerate}
\section{Motivation and Overview}
\label{sec:motivation}

\subsection{A Brief Introduction to Filter Lists}
\label{sec:motivation:filter-lists}
Filter lists are collections of regular-expression like rules describing trust statements
over URLs. The most common applications of filter lists are in browsers and extensions
to block unwanted requests when browsing the \Web{} (\EG{} requests for tracker, unwanted
advertisements, distracting page content, etc). Usually filter list rules describe
origins and paths that should be blocked, but most tools that apply filter lists
have additional syntax to further restrict how and when each rule should be applied.
For example, rules can be restricted to only be applied to certain kinds of requests (\EG{}
images, sub-documents, scripts) or only applied in certain contexts (\EG{} specifying
that some rules should only be considered when visiting certain sites).

Most popular filter lists are crowd sourced by communities that add and refine rules to
common lists. Rules are added when a list contributor finds out about a new
tracking script (or otherwise unwanted \Web{} resource) and decides to block it.

For example, assume a filter list maintainer is browsing a site and notices the
site has included a tracking script, served from \underline{https://tracker.example/bad.js}.
The filter list maintainer, wanting to protect other users, adds a new rule
to the filter list, instructing the browser to block the tracker on the current
site\footnote{This rule might look like
\texttt{\underline{||tracker.example/bad.js\$domain=}} \texttt{\underline{site.example}.}}.

The filter list maintainer then tests out the rule by revisiting the site
while applying the new rule. The maintainer sees that the tracking script is now
blocked, and that the site continues working correctly.

Having checked that the filter list rule works (\IE{} the target script was blocked) and
that the site still works, the filter list maintainer commits the new rule
to the filter list, which is soon downloaded by millions of filter list subscribers, blocking
the tracking script on that specific site for all users.

\subsection{Breaking Sites is Too Easy}
Filter list maintainers, though, have to choose between privacy and compatibility. Worse, they
often have to try and choose between these goals without data, and relying only on intuition and
best guesses.

To see why, refer back to the example discussed in the previous sub-section. The filter list
author specified that the tracking script should only be blocked when it is included by
one specific site; the tracking script will continue to be loaded on every other
site on the web, continue to harm users despite filter list author identifying the
script as a tracker. The privacy harm continues because the rule was written narrowly.

Alternatively the filter list author could have written the rule to be general, and to block the tracking
script whenever it was included on \emph{any} site\footnote{This general rule might look like
\texttt{\underline{||tracker.example/bad.js}}.}. This would prevent privacy harm, but risks
breaking sites. The filter list author only checked that one specific site still worked
when the script was blocked; other sites might have integrated
the script in such a way that the page breaks if the script is not present. This is common,
and happens when pages rely on utility functions tracking scripts provide (or otherwise deeply integrate
the tracking script).

In this scenario, not only has the filter list author broken an unknown number of pages,
but worse, the filter list author won't find out about the breakage until \emph{after the
rule has been shipped to users}, when users start encountering broken sites and (hopefully) reporting issues.

The underlying problem is that filter list maintainers have no scaleable, automated way to test rules before
shipping them. Maintainers can browse sites with the rule enabled, but this only works for rules
tied to a small number of sites; it does not scale to real rules that impact huge fractions of
the Web. The \Web{} is too large, and the number of filter list maintainers too small.

\subsection{Towards Automated Detection of Breakage}
Filter list authors need tools to help them protect user privacy, while minimizing risks to compatibility.
The ideal solution would be an oracle that allowed filter list authors to submit a proposed
filter list rule, and receive back a list of sites the rule would break. Further, this system
should be automated, so that filter list authors can repeatedly and quickly query it,
so that rules can be optimized (\IE{} maximizing privacy while minimising breakage) before shipping them to
users.

In practice this is difficult. Determining if a site is broken is tricky for a variety of reasons.
The broken functionality might not be immediately obvious, and might only be triggered after
interacting with the page. Blocking a script on one site might not effect the users at all,
while blocking the same script on another site might break the site entirely. Or, breaking a page
might not have any visual side effect at all; breakage might only manifest itself through unintended
application flow. These are just some examples of why ``site breakage'' is a difficult classification problem.

However, as a step towards building an automated site breakage oracle, we designed a system that
predicts whether a website will break, given three inputs: i) a webpage (described by URL),
ii) a filter list rule, and (optionally) iii) a browser profile, allowing the browser to be
arbitrarily configured before classification.

We developed our system using a ground truth dataset constructed from
the commit history of the EasyList project, and consisting of tuples of i) a \Web{} page URL, ii) a filter
list rule, and iii) whether the filter list rule broke the site (Section
\ref{sec:data-set:collecting}). We visited each URL in a crawler instrumented
to records the page's execution at an extremely detailed level (Section \ref{sec:dataset:capture}).
We then extracted \NumFeatures{} features from the execution record for each site, and used those features to
train a classifier (Section \ref{sec:classification_pipeline}) that performs with mean AUC
of \ClassifierMeanAUC{} (Section \ref{sec:results:utility}). Finally, we used the
classifier to learn \NumImportantFeatures{} features that predict whether a filter
list rule will break a website (Section \ref{sec:results:feature-importance}).
\section{Dataset}
\label{sec:dataset}

\begin{figure}[!t]
\begin{center}
\begin{tikzpicture}[node distance=1.1cm, align=center]
  \node (user0)     [user,draw=white,fill=white]
    {EasyList commit history};

  \node (sect1)     [sect, below of=user0]
    {\bf Parse, filter, and transform commits (\S\ref{sec:data-set:collecting})};

  \node (result0)     [result, below of=sect1, yshift=-0.3cm]
    {Tuples of \{page URL, filter list diff, broken/working label\}};

  \node (sect2)     [sect, below of=sect1, yshift=-1.7cm]
    {\bf Generate pre-/post-intervention filter lists (\S\ref{sec:dataset:capture})};
      

  \node (sect3)     [sect, below of=sect2, yshift=-0.3cm]
    {\bf Capture page behavior with \PG{} (\S\ref{sec:dataset:capture})};


    
  \node (sect5)     [sect, below of=sect3, yshift=-0.3cm]
    {\bf Post-process graph data (\S\ref{sec:data-set:post})};

  \node (result3)     [result,below of=sect5, yshift=-0.3cm]
    {Tuples of \{page URL, filter list diff, broken/working\\label, page behavior recording graphs\}};

  \draw[line width=1.5pt,->] (user0) -- (sect1);
  \draw[line width=1.5pt,->] (sect1.south) -- (result0.north);
  \draw[line width=1.5pt,->] (result0.south) -- (sect2.north);
  \draw[line width=1.5pt,->] (sect2.south) -- (sect3.north);
  \draw[line width=1.5pt,->] (sect3.south) -- (sect5.north);
  \draw[line width=1.5pt,->] (sect5.south) -- (result3.north);
\end{tikzpicture}
\end{center}
\vspace{3pt}
\caption{Pipeline diagram of our \Web{} compatibility dataset generation process.}
\label{figure:pipeline}
\end{figure}
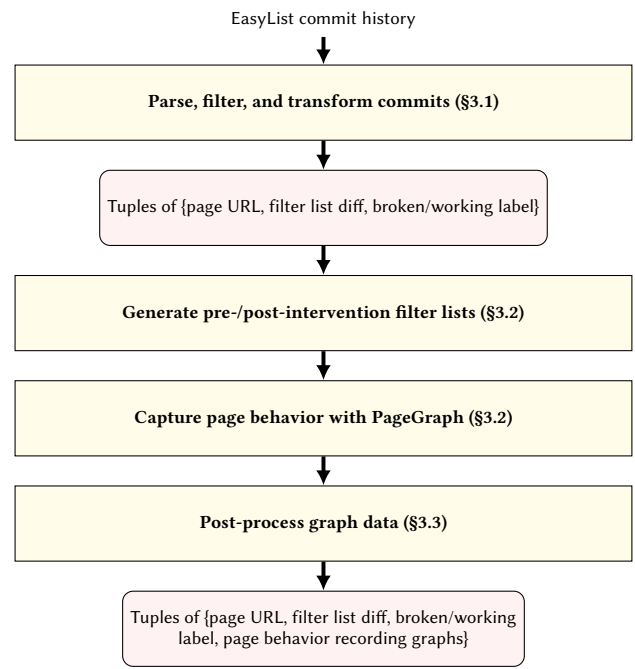
Our first contribution is a dataset of examples of filter list changes and their effects on page behavior, labeled with whether or not those effects represent \Web{} compatibility breakage from a user perspective, and the novel methodology with which we assembled this dataset at a sufficiently large scale for \ML{} classifier training.
The dataset contains a total of \NumDataPoints{} examples, each consisting of a page URL, a corresponding filter list change affecting the page, a broken-or-working label, and recordings of how the page's behavior responds to the change.
Figure~\ref{figure:pipeline} summarizes our data collection pipeline.

\subsection{Collecting Examples of Broken and Working Sites}
\label{sec:data-set:collecting}

To train our classifier to detect when a filter list change breaks a site, we first needed a set of examples for the classifier to learn from, of sites breaking when such a change is introduced.
Manually hunting through the \Web{} for broken sites, and then debugging filter lists to identify the rules responsible in each instance, would have been too time- and labor-intensive given the dataset size required to effectively train and test the classifier: our final dataset contains over a thousand such examples.
Moreover, this approach would place us as the judges of page brokenness, a subjective measure: our judgments may differ from those of end users.

\begin{figure}[!t]
  \begin{lstlisting}[
    style=code,
    linebackgroundcolor={
    \ifnum
        \value{lstnumber}=9
        \color{green!30}
    \else
        \color{editorGray}
    \fi}
    ]
P: https://www.mealty.ru/catalog/ (Fixes 
  https://forums.lanik.us/viewtopic.php?t=47335)
---

easyprivacy/easyprivacy_allowlist_international.txt:
  ...
  @@||mc.yandex.ru/metrika/tag.js$script,
    domain=auto.yandex|coddyschool.com
+ @@||mealty.ru/js/ga_events.js$~third-party
  @@||megafon.ru/static/?files=*/tealeaf.js
  ...
  \end{lstlisting}
  \caption{A sample \Web{} compatibility fix commit excerpted from the EasyList repository\protect\footnotemark, inserting an exception rule to allow through a script which shares its filename with a popular analytics script. Blocking the script breaks a page on \texttt{\underline{mealty.ru}}.}
  \label{figure:easylist_commit_p}
\end{figure}

\footnotetext{\url{https://github.com/easylist/easylist/commit/a509c21b72c2d4959bff05394082821f207730fd}}

\begin{figure}[!t]
  \begin{lstlisting}[
    style=code,
    linebackgroundcolor={
    \ifnum
        \value{lstnumber}=8
        \color{green!30}
    \else
        \color{editorGray}
    \fi}
    ]
A: https://tinyzonetv.to/
Block adserver at https://tinyzonetv.to/
---

easylist/easylist_adservers.txt:
  ...
  ||sftapi.com^
+ ||sfzover.com^
  ||sg2rgnza7k9t.com^
  ...
  \end{lstlisting}
  \caption{A sample coverage-expanding commit excerpted from the EasyList repository\protect\footnotemark, inserting a rule to block an ad server at \texttt{\underline{sfzover.com}}, found on the site \texttt{\underline{tinyzonetv.to}}.}
  \label{figure:easylist_commit_a}
\end{figure}

\footnotetext{\url{https://github.com/easylist/easylist/commit/0c453dbe0882640ce16dc823fc72dc3aaa55ec62}}

We sidestep these problems by mining labeled examples of \Web{} page breakage from a non-traditional source: the commit logs of the \EL{} project\footnote{\url{https://easylist.to/}}, a large and widely-used community-maintained filter list distribution.
The \EL{} authors use the Git version control system to coordinate the development of their filter lists, so each update to the rules is logged with an associated commit message, numbering over 169,000 across the project's history.
The commit messages follow uniform conventions agreed on by the authors.
In particular, a rule update to fix \Web{} compatibility breakage should be tagged with the prefix ``P:'' and reference the URL of at least one page on which the problem occurs.
Figure~\ref{figure:easylist_commit_p} shows a sample compatibility-fix commit taken from \EL{}.
These commits are often made in response to user reports, and are further vetted by the domain-expert maintainers that merge them into the \EL{} repository; therefore, we claim that they represent something close to ground truth for page breakage as perceived by end users.
Each of these commits typically comprises one or a few rule additions and/or deletions, constituting a filter list change which repairs compatibility breakage on the referenced page.
Inverting the change---i.e., flipping additions to deletions and deletions to additions---produces a filter list change which \emph{breaks} the page instead of repairing it.
By scanning the \EL{} commit history for \Web{} compatibility commits, parsing out the associated URLs, and applying this inversion to each commit, we seeded our dataset generation with \emph{positive} examples of filter list changes that introduce breakage, tied to specific \Web{} pages on which that breakage occurs.

In order to teach our classifier to distinguish filter list changes which break pages, we also needed to show it examples of changes which \emph{don't} cause breakage.
These changes should still have an effect on their target pages, but a desirable one: blocking an ad, for example, or circumventing a privacy-invading tracker.
Again we returned to the \EL{} commit logs, where such changes are tagged with the prefix ``A:'', and also reference page URLs on which the intended effects can be observed.
Figure~\ref{figure:easylist_commit_a} shows a sample from the \EL{} Git repository.
We applied the same process of scanning the commit history, minus the inversion step, to seed our dataset generation with \emph{negative} examples: non-breaking filter list changes and specific \Web{} pages they affect.

\subsection{Capturing Page Behavior}
\label{sec:dataset:capture}

\begin{figure}[!t]
\begin{center}
\begin{tikzpicture}[node distance=1.5cm, align=center]
  \node (img)        [base, dom, align=flush left]
    {{\bf DOM node}\\\texttt{\footnotesize id=197, tag=``img'',}\\\texttt{\footnotesize src=``https://a.com/b.png''}};
  \node (parser)     [base, parser, above of=img, yshift=0.6cm, align=flush left]
    {{\bf parser}\\\texttt{\footnotesize id=1}};
  \node (resource)   [base, resource, below of=img, xshift=4.4cm, yshift=-1cm, align=flush left]   
    {{\bf network resource}\\\texttt{\footnotesize id=218, type=``image'',}\\\texttt{\footnotesize url=``https://a.com/b.png''}};

  \draw[line width=1.5pt,->]
    (parser) --
    node [right, yshift=0.1cm] {{\bf \footnotesize node create}}
    (img);
  \draw[line width=1.5pt,->]     (img) |-
    node [above, xshift=1.2cm]
      {{\bf \footnotesize HTTP request}}
    node [below, xshift=1.4cm, text width=1in, align=flush left] 
      {\texttt{\footnotesize type=``Image'', size=1880, headers=``...''}}
    (resource);
  \draw[line width=1.5pt,->]     (resource) |-
    node [above, xshift=-1.0cm]
      {{\bf \footnotesize HTTP response}}
    node [below, xshift=-0.9cm, text width=1in, align=flush left] 
      {\texttt{\footnotesize status=200, size=13191, headers=``...''}}
    (img);
\end{tikzpicture}
\end{center}
\caption{\PG{} encoding of the initialization of an image element. The browser's HTML parser creates a DOM node to represent a decoded \texttt{<img>} tag. An HTTP request is dispatched to retrieve the image file pointed to by the image element's \texttt{src} attribute, and a success response is received.}
\label{figure:page_graph_sample}
\end{figure}
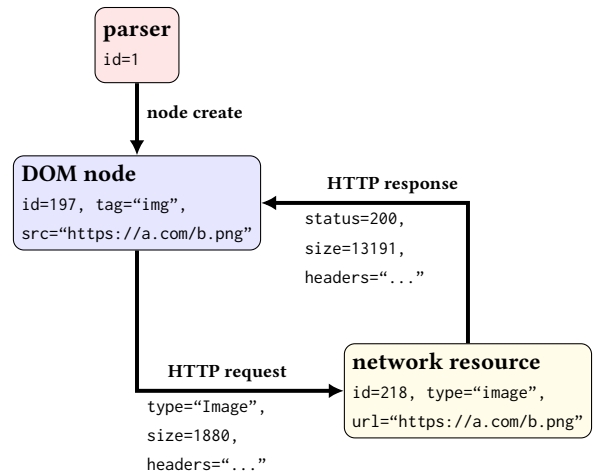

It would be difficult if not impossible to predict page breakage by looking just at \Web{} page URLs and filter rule source code.
Instead, our classifier draws its predictions from the way those pages behave at runtime in a browser equipped with content blocking, and how their behavior is altered by changes to the content blocker's filter lists.
We recorded page behavior with \PG{}, our deep browser instrumentation system built into the \BB{}.
Every page opened in a \PG{}-enabled browser build is monitored at runtime, and its activity across the browser engine is recorded in a unified directed graph structure.
Nodes of this graph correspond to interacting entities in the mini-ecosystem of a \Web{} page: actors like scripts, the parser, and the content blocker which perform actions, and those that are acted upon, like network resources, DOM nodes, \Web{} APIs, and filter rules.
Edges represent the actions that connect them: \texttt{node insert} rules between the parser and DOM nodes, \texttt{resource block} edges between filter rules and network resources, \texttt{API call} edges between script actors and \Web{} APIs; as well as \texttt{structure} edges which record parent-child relationships between DOM nodes.
Figure~\ref{figure:page_graph_sample} shows a sample excerpt from a \PG{} graph, illustrating how a \Web{} request to load the image pointed to by the URL of an HTML \texttt{<img>} tag is encoded as a pair of \texttt{HTTP request} and \texttt{HTTP response} edges, connecting the DOM node for the \texttt{<img>} tag to the \texttt{network resource} node representing the image file at the given URL, all annotated with metadata captured from this runtime interaction.

From the collection of positive and negative examples assembled in Section~\ref{sec:data-set:collecting}, we used our \PG{}-enabled \BB{} build to crawl each \Web{} page twice, capturing its behavior both \emph{with} and \emph{without} the corresponding filter list change (the ``intervention'') applied.
First, we used the \EL{} Git repository to generate a version of the filter list without the change.
We then injected this filter list into the browser's content blocker, replacing the built-in filter list.
Our crawler launched the browser, controlling it with the Puppeteer automation framework\footnote{\url{https://puppeteer.github.io/puppeteer/}}, and navigated to the page URL.
The crawler waited for the DOM \texttt{load} event to fire, and a further 15 seconds beyond that to give the page time to fully settle.
The page's behavior during this time was captured by \PG{} and exported in graph form at the end of the browsing session, taking a baseline for the page's normal operation; we refer to this as a \emph{pre-intervention} graph.
Next, we generated a second version of the filter list, this time with the change applied, and injected it into the content blocker.
Repeating the crawling process produced a second graph, representing the page's behavior under the influence of the filter list change; we refer to this as a \emph{post-intervention} graph.
For positive examples, this graph reflects the page in a broken state; for negative examples, it reflects a desirable, intentional change in behavior (e.g., blocking an ad).

Because we were using historical data---the \EL{} commit log---to drive our crawling, there was a chance that the pages we were visiting had been altered or had even disappeared in the time since the original commits were made.
As an initial heuristic, we only considered commits dated 2013 or later.
We rejected any pages which produced an error response (e.g., 404 Not Found) during crawling.
We further validated that applying the corresponding filter list change still actually had an effect on the page, by using the \texttt{adblock-rust}\footnote{\url{https://github.com/brave/adblock-rust}} library to evaluate the filter rules with and without the change against the page's recorded network activity.
If there was no difference between the blocking decisions made in the two cases, the page was excluded from our dataset.

\subsection{Post-Processing Data}
\label{sec:data-set:post}

At this point, we had two recorded \PG{} graphs for each example, one representing the page's behavior without the corresponding filter list change applied (``pre-intervention''), and one representing its behavior with the change in place (``post-intervention'').
We post-processed this dataset to generate a third ``intervention-only'' graph per example, which approximated the delta in page behavior caused by the intervention.
Each intervention-only graph is a sub-graph of the pre-intervention graph, containing graph nodes and edges missing or altered in the post-intervention graph.
We hypothesized that providing these narrowed-down sub-graphs as part of the input to our classifier would help it find a stronger signal, tuning out some of the noise of surrounding page behavior unaffected by the filter list change.

To generate an intervention-only graph, we first identify \texttt{network resource} nodes in the pre-intervention graph that the content blocker allowed through, but which are subsequently marked as blocked in the post-intervention graph.
These nodes represent the network resources covered by the filter rule change.
Starting from these resource nodes, we selectively walk outward in the pre-intervention graph, marking additional nodes and edges for inclusion in the sub-graph.
We walk up from each resource node to the node which caused the resource to be requested over the network, e.g., from image resource nodes to the HTML \texttt{<img>} DOM nodes pointing to those images; we mark these nodes and the connecting \texttt{HTTP request/response} edges for inclusion.
For any HTML \texttt{<script>} DOM nodes we discover, we follow \texttt{script execute} edges leading out from them to \texttt{script actor} nodes that represent those scripts running in the browser's JavaScript engine.
Finally, we mark all additional nodes which are reachable by taking one step from any already-marked node, as well as the connecting edges.
This captures, e.g., the effects that scripts blocked by the filter list change have on the page when not blocked, like the insertion or modification of DOM nodes, the registration and un-registration of event listeners, and \Web{} API calls.
Extracting the marked nodes and edges produces a new graph focused on the effects of the intervention on page behavior.
\section{Classifier Construction}
\label{sec:classifier}
We aim to construct a machine learning (\ML{}) classifier with sufficient accuracy to predict whether a filter list rule will break a site. 
Additionally, we aim to analyze each classifier's predictions to better understand which features predict site breakage. As a simple baseline, we employed a classical feature-based \ML{} model over an end-to-end learning system such as deep neural networks. The learning task is posed as a binary classification problem with the positive label \emph{``site did break''} and the negative label \emph{``site did not break''}.

In this sections we outline the data pre-processing and feature extraction, followed by a description of the full classification pipeline.

\subsection{Data Pre-Processing}
Our dataset numbers \NumDataPoints{} examples, each consisting of a ``pre-intervention'' graph, a ``post-intervention'' graph, and an ``intervention-only'' graph (as defined in Section \ref{sec:data-set:post}).
We excluded any examples with empty intervention-only graphs (indicating no measured effect on the page resulting from the intervention).
This left \NumTrainingSamples{} training samples with \NumPositiveLabels{} positive labels and \NumNegativeLabels{} negative labels.
As a pre-processing step, we converted each graph into pandas\footnote{\url{https://pandas.pydata.org/}} data frames, with each graph represented as an edge list.

\subsection{Generating Candidate Features}
\label{sec:feature_extraction}

We next describe how we generated the set of candidate features considered when constructing our classifier (Section \ref{sec:classification_pipeline} describes how we determined which features had significant predictive value, and Section \ref{sec:results:feature-importance} presents which features ended up being predictive).

To generate the candidate features, we first defined three dimensions that might be useful for predicting site breakage; each dimension is described below.
We then generated \NumFeatures{} features by selecting different options from each feature dimension.

\subsubsection{Scope of Analysis}
\label{sec:classifier:features:scope}
The first dimension we considered was whether to extract features
from i) the behavior of the overall page, or ii) the behavior that was blocked
by the the filter list rule.

\textbf{Page scope} features capture whether patterns in a pages' overall
behavior predicts breakage. These features would be predictive 
if certain aspects of the page's design and implementation predicted breakage,
independent of what was blocked on the page.
For example, ``page scope'' features might detect if site
complexity \emph{in general} predicts breakage. We extracted ``page scope'' features
from the ``pre-intervention'' graph (as described in Section \ref{sec:data-set:post}).

Conversely, \textbf{intervention scope} features look for patterns that
predict breakage specifically in the page behaviors blocked or modified by the
filter list rule. These features will be predictive if what is being blocked
predicts breakage (instead of the page context that blocking is occurring
in). ``Intervention scope'' features would be predictive if, for example, blocking
certain \JS{} API calls causes pages to break. We extracted ``intervention scope'' features
from the ``intervention-only'' graph (see Section \ref{sec:data-set:post}).

\subsubsection{Absolute vs. Relative Values}
\label{sec:classifier:features:count}
The second feature dimension we considered was whether to quantify behaviors using
i) absolute counts, or ii) as relative ratios.

Features using \textbf{absolute counts} are based on the number of times an event or element was observed
during a page's execution.
For example, an ``absolute count'' feature would be based on the number of video elements that were embedded in a page, or the number of network calls that were blocked by a filter list. ``Absolute count'' features could be used to detect if the size of a page, or the number of images on a page, can predicted breakage, independent of how many elements or images were blocked.

Features targeting \textbf{relative counts}, on the other hand, capture what percentage of
occurrences of an event on a page were blocked by the filter list rule.
For example, a ``relative count'' feature would consider the percentage of images blocked on a page, or the number of network requests prevented because of the filter list rule.

\subsubsection{Expertly Curated vs. Automatically Generated Target Behaviors}
\label{sec:classifier:features:generation}
The third feature dimension we considered was what strategy we used to decide
what page behaviors to measure. Some of the behaviors we targeted
were manually curated, drawing from domain knowledge and
``expert'' intuition; other features were automatically generated by examining
generic graph features.

The \textbf{expert curated} features were generated by extrapolating
from our domain knowledge and past experience dealing with broken websites. Our
domain knowledge comes form various sources, including our research and experiences contributing to and maintaining privacy tools\footnote{\PrivacyContributionsLinks{}}. We then generalized
our observations about how websites break, and tried to capture those generalizations
as features. Some examples of ``expert curated'' features we generated include
i) whether a blocked script fetches additional scripts, or
ii) whether a blocked script registered event handlers in the page.

We also \textbf{automatically generated} a large number of additional features
that considered counts of different graph attributes (node types, edge types, node attributes,
edge attributes). These automatically generated features did not consider the relationship
between different graph elements, or the semantics of the values for node and edge attributes.
As a result, the ``automatically generated'' features were generally much simpler
than the kinds of page behaviors captured in the ``expert curated'' features.
Some examples of ``automatically generated'' features include
i) how many DOM nodes of each HTML tag appeared in the page (HTML tag names are recorded in node attributes in \PG{}), or ii) how many Web APIs were called during a page's execution (Web API invocations are
recorded in \PG{} with edges of type ``call'').

We then categorized our features (whether
``expert curated'' or ``automatically generated'') into one of the following five
categories, depending on the kinds of page behaviors were measured by the feature.

(see Section \ref{sec:results:feature-importance}).

\begin{itemize}
    \item \textbf{HTML structure}: aspects of the structure and composition of the document
        (\EG{} numbers of different tags, amount of text on the page)

    \item \textbf{\JS{} modifications of page structure}: measures of how the page's structure
        was constructed or modified by scripts (\EG{} numbers of DOM nodes inserted by scripts,
        amount of text modified by scripts)

    \item \textbf{Other \JS{} behaviors}: script operations and calls not directly related to
        modifying DOM structure (\EG{} counts of API calls made by scripts, numbers of
        event handlers registered by scripts)
        
    \item \textbf{Network behaviors}: the sub-resources and other network calls made
        during page execution (\EG{} number of bytes fetched, number of sub-resources fetched)
    
    \item \textbf{Generic graph features}: graph measurements with considering the behaviors
        being encoded by the graph (\EG{} number of nodes and edges, number unique node attribute
        values)
    
\end{itemize}

We note that this categorization was not used in the training or evaluating of the classifier;
this categorization was instead to benefit the later discussion of what kinds of
page behaviors predicted breakage (Section \ref{sec:results:feature-importance}).


\subsection{Classification Pipeline}
\label{sec:classification_pipeline}
\label{sec:classifier:pipeline}

The classification pipeline consists of the model to learn the target function based on the extracted features as described in the previous section, as well as several steps to transform the inputs before making predictions.

We select \XGBOOST{}\footnote{https://github.com/dmlc/xgboost}, a popular model choice which has been shown to achieve state-of-the-art performance across a wide range of prediction tasks \cite{chen2016xgboost}. As a tree-based method, \XGBOOST{} has several characteristics that make it particularly suitable to serve as an off-the-shelf baseline method. 
First, variable selection is performed automatically, making it immune to the inclusion of irrelevant features. 
We empirically verify this by conducting recursive feature selection over all \NumFeatures{} input features, and found that it has no significant effect on performance. Second, tree-based methods are robust to outliers due to the way they partition the input space. We verify this empirically by removing the top 1-percentile of training samples w.r.t. their node/edge ratio, finding no significant effect on performance. Last, tree-based methods naturally deal with missing values (see e.g. \cite{hastie2009elements, murphy2012machine}).

Despite the ability for tree-based methods to handle missing values and correlated features, we conduct the following transformations on the input data in order to improve robustness of the model and reduce training time:
\begin{enumerate}
    \item All features which contain a percentage of empty values above a certain threshold are dropped (empirically we found a good threshold to be \CorrelationThreshold{}, see section \ref{sec:evaluation1}).
    \item All features which have a Pearson correlation coefficient above a certain threshold with at least one other feature, are dropped (empirically we found a good threshold to be \NullValueThreshold{}, see section \ref{sec:evaluation1}).
    \item All remaining features are standardised by removing the mean and scaling to unit variance.
\end{enumerate}

\section{Classifier Evaluation}
\label{sec:evaluation_and_results}
This section outlines the analysis of the classifier described in the previous section with respect to its predictive and explanatory power, as well as its sample complexity. We divide the evaluation into three parts:
\begin{enumerate}
    \item In section \ref{sec:evaluation1} we train, tune and test the classifier in order to optimise its predictive performance and show whether it is possible to achieve practical utility on the \PG{} data set.
    \item In section \ref{sec:results:feature-importance} we train and test the classifier (without tuning) on subsets of features to analyse the effect of individual features on the predictions.
    \item In section \ref{sec:evaluation3} we train and test the classifier (without tuning) on training data samples of increasing size, to understand how much data is needed to achieve practical utility.
\end{enumerate}
To summarise and evaluate the accuracy of the predicted class probabilities, we use the area under the receiver operating characteristic curve (\ROCAUC{}) as a threshold-invariant metric.

\subsection{Practical Utility}
\label{sec:evaluation1}
\label{sec:results:utility}
First, we evaluate the performance of the classifier with respect to its practical utility. To that end, we conduct several hyper-parameter optimization rounds via nested cross-validation to optimize the predictive performance of the classifier. In the inner loop, the validation fold is rotated across \NumInnerFolds{} folds to choose the best hyper-parameter configuration (modelled as samples from a \textit{Gaussian Process} using the framework of \textit{Bayesian Optimization} with \NumOptimisationSteps{} parameter configurations being tested per fold). Hereby, we optimize over all pipeline steps including the thresholds for dropping null-valued and correlated features. The outer loop with \NumOuterFolds{} folds is used to evaluate the performance of the learner. The evaluation is conducted over all \NumTrainingSamples{} training samples with all \NumFeatures{} features. 

\begin{figure}[t]
    \centering
    \includegraphics[width=\columnwidth]{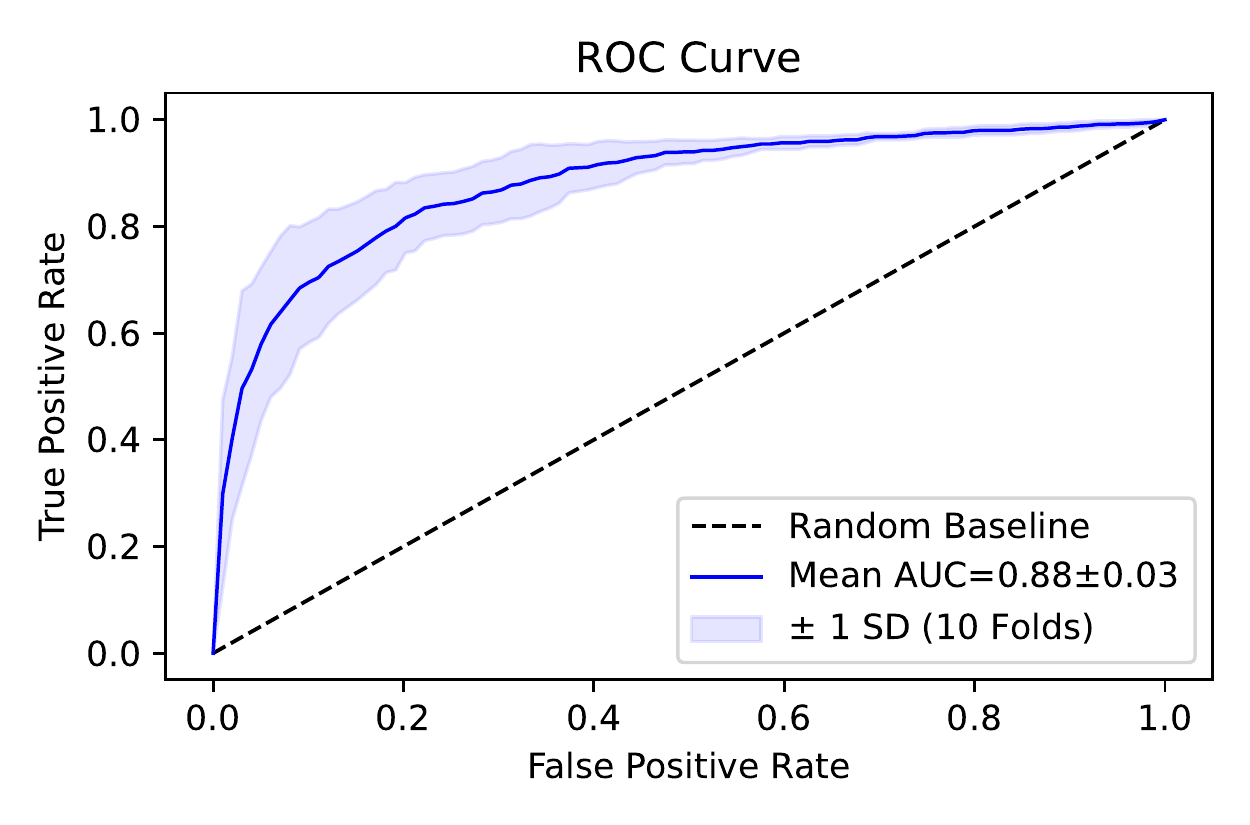}
    \caption{\ROCAUC{} curves averaged over \NumOuterFolds{} folds with hyper-parameter tuning to evaluate the practical utility of the classifier.}
    \label{figure:roc_curve}
\end{figure}

As shown in Figure \ref{figure:roc_curve}, we find that the classifier can achieve practical utility, with an average \ROCAUC{} of \ClassifierMeanAUC{}($\pm$\ClassifierStdAUC{}) across the outer \NumOuterFolds{} folds. We thus show that an off-the-shelf classifier can extract enough signal from the \PG{} data set to separate positive examples (breaking) from negative examples (non-breaking) well above random. Depending on the modelling objective it is possible to further optimize performance with respect to precision and recall or type 1 and type 2 error by tuning the classification threshold.

\subsection{Feature Importance}
\label{sec:results:feature-importance}

Next, we analyze the explanatory power of the classifier by inspecting the effect of the extracted features on predictions. First we conduct the analysis along the lines of the feature generating dimensions by removing all features belonging to a specific dimension and measuring the drop in predictive performance. We then repeat the analysis for individual features to establish a rank-order of the most predictive features. Despite the fact that the \textit{Leave-One-Covariate-Out}\cite{lei2018distribution} importance metric doesn't capture interaction effects if applied to single features, it can nonetheless hint at the relative amount of signal contained in individual features.

To estimate the importance metric for feature groups and single features respectively, we first fit the classifier with default parameters (no tuning) and all \NumFeatures{} features to serve as a baseline. We then generate a new data set for each feature (group) by removing the respective features from the training data before fitting the model with the remaining features. We then estimate the importance by subtracting the \ROCAUC{} of the reduced feature set from the baseline for each feature respectively via \NumFeatureImpFolds{}-fold cross validation and report the mean and standard deviation.

\begin{figure*}
    \centering
    \includegraphics[width=\linewidth]{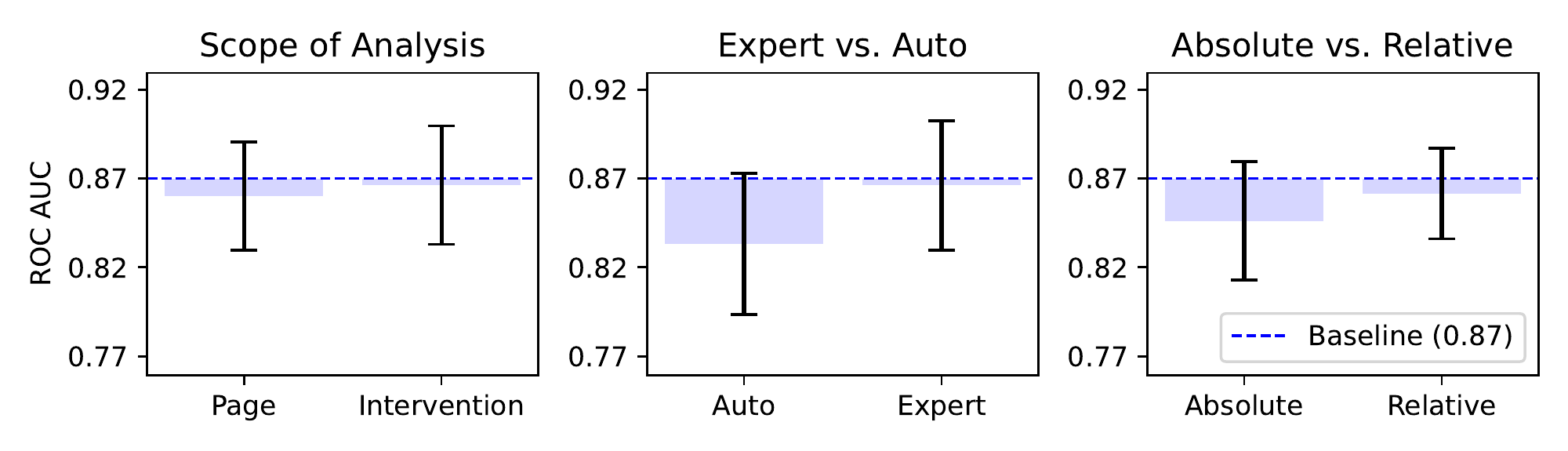}
    \caption{Importance of the feature generating dimensions \textbf{\nameref{sec:classifier:features:scope}}, \textbf{\nameref{sec:classifier:features:count}} and \textbf{\nameref{sec:classifier:features:generation}} as measured by the mean loss in \ROCAUC{} when removing the respective features from the training data. Error bars indicate the standard deviation across \NumFeatureImpFolds{} folds.}
    \label{figure:feature_imp}
\end{figure*}

The results for the importance of feature groups per feature generating dimension are shown in Figure \ref{figure:feature_imp}. Page features in isolation result in a slightly higher loss than Intervention features, with a mean loss of \ROCAUC{} \MeanLossAUCForPage{} and \MeanLossAUCForIntervention{} respectively. More pronounced is the difference between expertly curated vs. automatically generated features with the auto-generated features incurring a mean loss of \ROCAUC{} \MeanLossAUCForExpert{} and  \MeanLossAUCForAuto{} respectively. Slightly less pronounced is the difference between absolute and relative features with a mean loss of \ROCAUC{} of \MeanLossAUCForAbsolute{} and \MeanLossAUCForRelative{} respectively.

Results for the importance of individual features are shown in Table \ref{table:feature-importance}, and are described in detail in the following section.

\subsubsection{Features with Predictive Power}
\label{sec:results:features:predictive}

\begin{table*}[!ht]
  \center
  \begin{tabular}{rllrl}
      \toprule
          Rank    & Scope & Source & AUC Loss    & Description \\
      \midrule
          \multicolumn{5}{l}{Page Structure Features} \\
      \midrule
          3 & Intervention & Auto & 0.00374 & \% of sub-document requests blocked \\	
          6 & Page & Auto & 0.00222 & \# of tags and text nodes in initial HTML \\	
          13 & Intervention & Auto & 0.00156 & $\Delta$ in \# of sub-documents after blocking \\	
          19 & Page & Auto & 0.00078 & \# of \texttt{<iframe>} in page \\	
          33 & Page & Auto & 0.00025 & \% of DOM nodes that are \texttt{<html>} \\	
          36 & Page & Auto & 0.00018 & \% of DOM nodes that are \texttt{<iframe>} \\	
          40 & Intervention & Expert & 0.00012 & \% of <html> elements blocked \\	
      \midrule
          \multicolumn{5}{l}{Generic Graph Features} \\
      \midrule
          11 & Page & Auto & 0.00179 & \# of unique node and edge types \\	
          22 & Intervention & Auto & 0.00061 & \# of unique types of actions taken by blocked scripts \\	
          35 & Page & Auto & 0.00023 & \# of unique types of actions in entire page \\	
      \midrule
          \multicolumn{5}{l}{Features Regarding \JS{} Modifying Page Structure} \\
      \midrule    
          7 & Intervention & Auto    & 0.00217 & \# of DOM nodes created by HTML parser prevented by blocking \\	
          12 & Intervention & Auto   & 0.00169 & \% of JS DOM nodes created by blocked scripts \\	
          16 & Intervention & Auto   & 0.00116 & \# of DOM node insertions done by blocked scripts \\	
          21 & Intervention & Expert & 0.00062 & \# of <html> elements created by blocked scripts \\	
          25 & Intervention & Expert & 0.00046 & \# of DOM nodes created by blocked scripts \\	
          30 & Page & Auto & 0.00032 & \# of DOM nodes created by scripts in entire page \\	
          34 & Intervention & Auto & 0.00024 & \% of DOM nodes deletions done by blocked scripts \\	

      \midrule
          \multicolumn{5}{l}{Other \JS{} Features} \\
      \midrule
          4 & Page & Expert & 0.00295 & \# of times any script accessed properties on \texttt{window.navigator} \\	
          5 & Intervention & Auto & 0.00254 & \# of scripts fetched or eval'ed by blocked scripts \\	
          8 & Page & Expert & 0.00216 & \# of times any script deleted a value from \texttt{sessionStorage} \\	
          9 & Intervention & Expert & 0.00205 & \% of \texttt{document.cookie} sets occurring in blocked scripts \\	
          10 & Intervention & Auto & 0.00190 & \% of \texttt{localStorage} operations occurring in blocked scripts \\	
          14 & Page & Auto & 0.00126 & \# of scripts fetched or eval'ed in entire page \\	
          15 & Intervention & Expert & 0.00120 & \# of times blocked scripts read from \texttt{document.cookie} \\	
          17 & Page & Auto & 0.00113 & \# of \texttt{document.cookie} operations in entire page \\	
          18 & Intervention & Auto & 0.00083 & \% of \texttt{sessionStorage} operations done by blocked scripts \\	
          20 & Page & Expert & 0.00073 & \# of WebGL calls, over the entire page \\	
          23 & Intervention & Auto & 0.00059 & \# of Web API calls made by blocked scripts \\	
          26 & Intervention & Auto & 0.00042 & \% of \texttt{eventListener} removals done by blocked scripts \\	
          28 & Page & Auto & 0.00034 & \# of \texttt{eventListener} registrations in entire page \\	
          29 & Intervention & Expert & 0.00033 & \% of \texttt{window.navigator} reads made by blocked scripts \\	
          31 & Page & Auto   & 0.00027 & \# of <script> tags in page \\	
          37 & Page & Expert & 0.00017 & \# of \texttt{window.screen} reads over entire page \\	
          38 & Page & Auto   & 0.00016 & \# of cross-document script-reads in entire page \\	
          39 & Page & Expert & 0.00013 & \# of \texttt{localStorage} reads over entire page \\	

      \midrule
          \multicolumn{5}{l}{Network Features} \\
      \midrule
          1 & Intervention & Expert  & 0.00831 & $\Delta$ in bytes sent over network after blocking \\ 
          2 & Intervention & Expert  & 0.00550 & size of resources directly blocked \\ 
          24 & Intervention & Expert & 0.00059 & \# of resources blocked (direct or indirect) \\  
          27 & Page & Auto           & 0.00034 & \% of page actions that were network requests \\ 
          32 & Intervention & Expert & 0.00026 & \% of network resources that were blocked \\ 
      \bottomrule
  \end{tabular}
  \caption{This table presents the \NumImportantFeatures{} features (from the \NumFeatures{} features considered) that predicted page breakage. The ``Rank'' column gives the relative importance of each feature, with 1 being the most predictive, 40 the least. ``Scope'' describes whether the feature was extracted from the \emph{pre-intervention} graph, denoted ``Page'' or \emph{intervention-only} graph, denoted ``Intervention'' (Section \ref{sec:classifier:features:scope}). ``Source'' gives whether the feature was developed through expert curation or automatic generation (re Section \ref{sec:classifier:features:generation}.  ``AUC Loss'' gives how much predictive power was lost when the feature was removed (Section \ref{sec:classifier:pipeline}).  ``Description'' provides a terse description of the page behavior captured by the feature.}
  \label{table:feature-importance}
\end{table*}

This process yielded \NumImportantFeatures{} features, from an initial starting set of
\NumFeatures{} candidate features. 
Table \ref{table:feature-importance} gives the \NumImportantFeatures{} features that had predictive power in our model.
A wide range of features ended up being useful, representing a wide range of page behaviors (\EG{} \JS{} API calls,
structure and size of the DOM, amount and size of network calls). As depicted in Figures \ref{figure:feature_imp},
no single kind of page behavior dominated the predictive power of the classifier (though, as discussed later,
features capturing pages' network behaviors were somewhat more predictive than other categories of features).

Similarly, no other ``dimenson'' of features dominated the classifier's predictions.
Nearly as many predictive features measured overall page behaviors (20) as measured
just the activities blocked by the filter list rule (20).
Similar numbers of expert-curated features were predictive as automatically generated features (15 and 25, respectively).

However, there were some trends we observed in which page behaviors predicted page breakage. 
We here briefly note three interesting and surprising patterns we observed.

\SubPoint{First, the kinds and number of \JS{} features used on a page predicted page breakage}
\JS{} behaviors that were blocked (\IE{} the scripts that were blocked by the filter list rule) were generally more predictive than over all page behaviors (\IE{} blocked and not blocked scripts alike).
The number of scripts fetched and executed by blocked scripts, the number of cookies set by blocked scripts, and the number (and \%) of DOM nodes injected by blocked scripts were all predictive of breakage. 

These findings support an intuition that most \Web{} applications are highly modular, with the privacy threatening portions of each application being contained in a small portion of the overall implementation. That does not, though, mean that the privacy harming parts of many \Web{} applications can be cleanly severed from the overall application; just the opposite in fact. These findings support the idea that current applications are both modular and tightly coupled, and that blocking the privacy harming scripts often breaks the overall application. This conclusion is supported by other related work\cite{smith2021sugarcoat} that finds that filter-list based content blocking tools are insufficiently granular to effectively address many kinds of privacy harms on the \Web{}.

\SubPoint{Second, the number of sub-documents in a page predicts page breakage}
Many of behaviors that predict breakage relate to the number of sub-documents included on the page, both directly
(\EG{} ``\% of sub-document requests blocked'', rank 3, or ``$\delta$ of \# sub-documents after blocking'', rank 13)
and indirectly (``\# of \texttt{<html>} elements created by blocked scripts'', rank 21). 
Similarly, whether the same \texttt{<iframe>} element was used to load multiple sub-documents further predicted page breakage.
For example, ``$\delta$ in \# of sub-documents after blocking'' (rank 13) was predictive of breakage, independent of the ``\# of \texttt{<iframe>} on the page'' (rank 19).

\SubPoint{Third, network behaviors were highly predictive of breakage}
Though only five of the \NumImportantFeatures{} predictive features in our set
directly related to network requests made by the page, this group contains
both the first and second most predictive features (``$\delta$  in bytes sent over network after blocking'' and ``size of resources directly blocked'', respectively). The number and percent of network requests blocked were also, unsurprisingly, predictive of breakage (rank 24 and 32, respectively).

\subsubsection{Features without Predictive Power}
\label{sec:results:features:not-predictive}

Additionally, we briefly note some features we expected (\IE{} they were ``expert curated'') to predict page breakage, but which ended up not being predictive. 
We give below a list of features we expected to be predictive, but which were not.

For example, we expected the number of blocked event registrations (\IE{} the number of events that blocked scripts would have registered) to predict page breakage.  
We expected this on the intuition that at least some of these event registrations would have been important page behaviors (\EG{} form handlers, interactive page elements),
behaviors that would ``break'' the page if they were omitted. 
However, this proved not to be the case; the number of blocked event registrations did not predict breakage. 
We suspect this might be because most event registrations in blocked scripts end up not being core to page behavior, and are instead more likely to be related to user tracking or other undesirable (to the user) behaviors (\EG{} interaction ``heat maps'' and other fine-grained behavioral tracking).

Similarly, despite our expectations, the number of text nodes inserted by blocked
scripts did not predict page breakage. Our intuition was that if blocking a script removes a lot of text from the page, that script is likely important to the page. 
However, this turned out to not be the case; this feature was not predictive of breakage. One possibility is that unwanted scripts (\EG{} large advertisements, captions for video ads, etc.) end up also being responsible for a large amount of text, and so ``amount of text added to the page'' ends up not being useful for distinguishing blocking that breaks a page from blocking that leaves the page functioning well.

\subsection{Sample Complexity}
\label{sec:evaluation3}
Last, we empirically analyze the sample complexity of the classifier in order to understand the amount of training data needed to achieve practical utility.

\begin{figure}[t]
    \centering
    \includegraphics[width=\columnwidth]{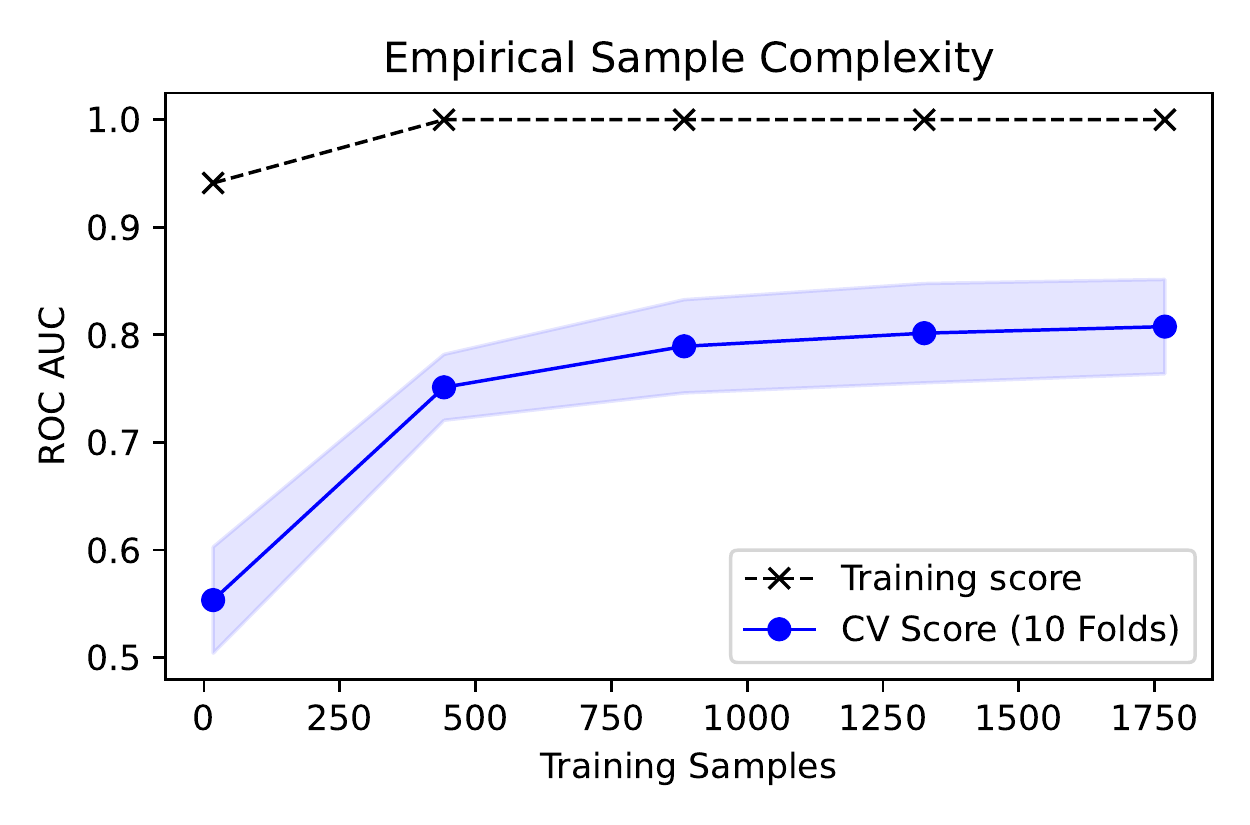}
    \caption{Empirical sample complexity estimated by the mean \ROCAUC{} as a function of the number of training samples across \NumOuterFolds{} folds.}
    \label{figure:sample_complexity}
\end{figure}

Analogous to the previous section we omit the tuning step and initialise the classifier with default parameters. 
We then train the classifier on varying amounts of data (1\%, 25\%, 50\%, 75\% and 100\% of training samples), each time reporting the mean \ROCAUC{} over \NumOuterFolds{} folds to estimate classification performance given the respective amount of data. 
Finally, we plot the mean \ROCAUC{} as a function of the number of training samples in Figure \ref{figure:sample_complexity}.

We can see that performance converges after training the classifier on roughly \ConvergedAfterPercentageSamples{}\% of the training data (\ConvergedAfterNumSamples{} samples), i.e. collecting and training on additional training samples yields diminishing returns. 
Given the cost of acquiring new samples, this is a desirable result.

\section{Discussion}
\label{sec:discussion}

We next provide some discussion of how our system could be deployed, extended
and improved by future work. We first discuss practical deployment scenarios
for our compatibility classifier, and how our system could be used by
filter list authors to improve the compatibility and coverage
of crowd sourced filter lists. Next, we discuss how our approach could
be extend to other, non-filter list privacy interventions, and the potential
difficulties in doing so. Last, we discuss some of the limitations and weaknesses
in our approach, and how they could be addressed by future work.

\subsection{Deployment Strategies}
\label{sec:discussion:deployment}

We here briefly discuss several ways our classifier could be used to
improve the usability of content blocking filter lists, and by site authors
to prevent the chance their sites break for users of content blocking tools.

\subsubsection{Aid Existing Filter List Auditors}
One possible deployment strategy for a web-compat classifier is to use the classifier
as an aid to existing human evaluators. In this case, the classifier could be tuned
to prefer recall over precision, and could be used to reduce the number of broken sites
a human evaluator would need to consider; the human labeler would still need to
distinguish broken cases from working cases, but the number of cases the human labeler
would need to consider would be reduced. The classifier would be, effectively, a force-multiplier
for existing filter list developers.

More concretely, the classifier could be used to crawl the \Web{} and look for potential breaking websites.
Filter list developers and maintainers would still be needed to distinguish the false positives from
true positives (optimizing for recall would reduce precision), but the number of cases
humans evaluators would need to consider would be significantly reduced, from ``the universe of all possible
websites'' to ``the false positives generated by the classifier.''

\subsubsection{Protecting Must-Work Sites}
Similarly, the classifier could be deployed to warn when a new rule might break ``high-priority'' sites,
as part a continuous-integration-style system for filter list development. Different communities of
filter list users (or developers) may have sites that are of extra-importance to
them (\EG{} high popularity sites, either globally or by linguistic community, or
sites vital to communication or safety with threatened groups, etc). In such cases, filter list
authors might wish to be extremely confident that these priority sites continue to work when new filter
list rules are added. Manually checking such sites every time a new filter list rule is added
would be prohibitive (popular filter lists are updated several times a day). An automated classifier
could reduce the amount of manual verification needed to manageable levels.

\subsubsection{Assisting Site Authors}
Finally, the classifier could also be used by site authors who wish to be warned when a new filter list
rule might break their site for users of content blocking tools. While we expect that most
site authors would prefer visitors not use content blocking tools at all, the popularity of
``please disable your content blocker'' notifications (often with a ``dismiss'' option for users)
suggest that a non-trivial number of site authors would prefer their sites work in the presence
of a content blocking tool, over their site breaking for the user all together. In these cases,
concerned site authors could use the classifier to receive an ``early warning'' when their
pages might break for filter list users.

Site hosting services, or reverse-proxy services (like Cloudflare or Fastly) could offer such
``breakage'' warnings as a service to their clients.

\subsection{Applicability to Other Privacy Interventions}
\label{sec:discussion:other-interventions}

Though we choose to build our classifier to predict when filter list rules could break a page,
we expect our approach could be extended to other privacy interventions. For example,
\Web{} privacy tools i) modify how third-party storage is managed, ii) use list or heuristic-based
approaches to protect users against navigational-tracking, and/or iii) deploy a range of
defenses against browser fingerprinting, among many other protections. All of these approaches
risk breaking websites, either because they fail to distinguish between benign and malicious
behavior, or because the malicious behavior they prevent is deeply entwined with desirable
page behaviors.

We expect that a classifier that could detect when these other privacy interventions break sites
would be beneficial to their developers and users, for many of the same reasons discussed in
Section \ref{sec:motivation}. Developers of these tools could use an automated classifier
to detect when, and what kinds of, sites break, and use that information to either refine
their tools, or create application-exceptions when needed.

While we expect the general approach of constructing and training our classifier would work
for other privacy interventions, extending our classifier to other systems would require
some non-trivial changes. For example, our work leverages the \EL{} commit history
to create a ``naturally occuring'' ground truth dataset; finding a comparable set of pre-labeled
data may be difficult for other projects. Similarly, many of the features we selected for predicting
page breakage are likely more applicable to filter-list blocking than other privacy interventions
(\EG{} changes in number of scripts requested, sub-documents loaded, or event handlers registered).
Detecting when other kinds of privacy interventions break pages likely will require other (or at least
additional) features.

\subsection{Limitations and Future Work}
\label{sec:discussion:limitiations}

Finally, we note some weaknesses and limitations in our approach, and suggest how they could be
improved through future work. First, our crawler does not interact with pages when recording (through \PG{})
page behaviors. As a result, there are likely cases where our crawler does not trigger some ``broken''
behaviors on the page, causing those broken behaviors to not be recorded in our dataset. This means
that our classifier is not considering some (possibly) predictive information, and so is not performing
optimally.

For example, consider the case when a filter list rule blocks a form validation script,
but that form validation script is only applied to the page after some user interaction (such as clicking on a
``contact us'' button on the page). Our crawler would record the script being blocked, but \emph{not}
how the page behaves when the user tries to submit the now-broken contact form. As a result, certain categories
of broken behaviors are missed by our crawler and classifier. Future work in this area might address
this weakness (and so likely improve the performance of the classifier) by having the crawler interact with
pages, either by having the instrumented browser be driven by a human user, or through software that
attempts to simulate human interactions\footnote{For example, stress-testing scripts like \url{https://github.com/marmelab/gremlins.js}.}.

Second, our system only considered the landing page of a site, even though its possible that
the ``broken'' part of the site only occurs on a child page in site. This limitation comes from our
selected ground truth dataset; as discussed in Section \ref{sec:dataset}, \EL{} commits only record
what sites are being fixed by a new commit, not what pages (\IE{} full URLs) are being fixed.
As a result, it is likely that some of the recording labeled as ``broken'' in our ground truth
are only displaying correctly-functioning behaviors (since the recorded page is not broken, some other
page on the same site is). This again suggests that our classifier is not optimal, and that performance
could be improved by finding (or creating) another dataset with more accurate labels.

Third, and more broadly, while our system can tell filter list authors when a site might be breaking,
our system does not provide the filter list author with an easy way of fixing the site. Filter list
authors could choose to remove the relevant rule, or modify the rule so that the rule is not applied
to the breaking site. This would maintain compatibility, but only by undermining the initial goal
of the filter list! Figuring out how to modify filter list rules so they protect privacy without
sacrificing compatibility is beyond the scope of work, but is an important area for future work. 

Fourth, we note that there are other kinds of features that could be used to possibly further
improve classifier performance. Our implementation only considered page behaviors when
trying to predict breakage, but our approach could be extended to consider other available
data. For example, features could be designed to consider visual differences in a
page before and after blocking, on the (possible) intuition that if applying a filter list
rule causes a large visual difference, its more likely the filter list rule has broken the page.
Future work could try boosting the classifier's performance with many such other sources
of information.
\section{Related Work}
\label{sec:rel-work}

In this section we discuss how our automated compatibility classifier compares and relates
to other work in the area, and specifically to existing research exploring how
the compatibility risks of privacy interventions, and the challenges and history of maintaining content blocking filter lists.

\subsection{Compatibility of Privacy Protections}
\label{sec:rel-work:compat}

Our work most directly relates to a focused but important area of research and practice around the compatibility costs of privacy enhancing techniques.

Much of the existing work in this area starts with proposed method of improving privacy for users, and then evaluating the compatibility costs of the proposed intervention.
For example, Yu \EtAl{}\cite{yu2016tracking} proposed an automated system for detecting trackers on the \Web{}, based on how often third-parties reoccurred across first-party sites. They then built an extension that
would block detected trackers, and then estimated how many websites their extension broke based on how often users reloaded pages. Snyder \EtAl{}\cite{snyder2017most} similarly suggested that \Web{} privacy and security
could be improved by removing infrequently used Web APIs, and had human labelers evaluate how many websites broke when each feature was blocked. 
Smith \EtAl{} proposed improving \Web{} privacy with an automated system
that would rewrite scripts to remove privacy harming behaviors without disrupting benign, user serving code paths.
They too evaluated their system with human labelers. 
Iqbal \EtAl{} 2020\cite{iqbal2020adgraph} and 2021\cite{iqbal2021fingerprinting}
used similar human evaluation systems for determining the compatibility impact of machine learning based approaches for blocking tracking scripts and detecting fingerprinting scripts, respectively.

Other research has focused on evaluating the compatibility trade-offs in existing \Web{} systems, instead of evaluating the compatibility of a newly proposed system. Jueckstock \EtAl{}\cite{jueckstock2022measuring}
used a human labeling system to evaluate how often different systems for managing third-party storage in \Web{} browsers broke sites. 
Mesbah \EtAl{}\cite{mesbah2011automated}, Choudhary \EtAl{}\cite{choudhary2011detecting}
and Van Deursen \EtAl{}\cite{van2015crawl} proposed systems for measuring when differences in browser implementations of \Web{} standards broke websites.

Finally, recent work by Mandalari \EtAl{}\cite{mandalari2021blocking} describes a system that determines when privacy interventions break user desirable systems, but for internet-of-things devices instead of websites. 
Their system automatically distinguishes necessary traffic flows from non-core
flows, and only applies privacy protections (\IE{} blocking) to traffic flows not necessary for a devices user-serving functionalities.

\subsection{Filter List Maintenance}
\label{sec:rel-work:filter-lists}

Our work also builds on a large body of work identifying and/or addressing difficulties in
maintaining content blocking filter lists. Snyder \EtAl{}\cite{snyder2020filters} measured how much
``dead weight'' (\IE{} non-useful rules) had accumulated in popular filter lists, and proposed
a system for optimizing filter lists by removing rules that were not ever applied during
automated crawls of the \Web{}. Chen \EtAl{}\cite{chen2021detecting} proposed a system for detecting
when trackers evade filter lists by moving, combining, or renaming tracking scripts by identifying scripts
by their behaviors (instead of their URLs). They proposed using their approach to automatically add ``evading''
scripts to existing filter lists. Sj{\"o}sten \EtAl{\cite{sjosten2020filter}} found that many region-specific
filter lists were not as well maintained as filter lists targeting languages with more global speakers (\EG{} English, Spanish, Chinese, etc),
and proposed a machine-learning approach for augmenting regional filter lists based on what is blocked by global lists. Bhagavatula \EtAl{}\cite{bhagavatula2014leveraging} proposed a system for assisting filter
list authors by using machine learning to detect textual patterns in blocked URLs, and to use that classifier to generate new filter list rules.

Alrizah \EtAl{}\cite{alrizah2019errors} measured how often, and how long it took for blocked scripts to
try and evade being blocked by filter lists, and found that it often takes filter list authors over a month to respond to
evasion efforts. Wang \EtAl{}\cite{wang2016webranz} similarly found filter lists have difficulty keeping up
when websites attempt to evade detection, though their work focused on websites modifying page structure to
avoid cosmetic filtering rules.

Finally, a body of work has studied the difficulties filter list authors face when sites attempt to block
filter list users (\EG{} ``anti-ad-block'', or ``ad-block-blockers''). Iqbal \EtAl{}\cite{iqbal2017ad} and
Nithyanand \EtAl{}\cite{nithyanand2016adblocking}, for example, both find that many sites attempt to detect when a visitor is applying a filter list (either by checking for blocked requests or for hidden page elements)
and apply a range of countermeasures to try and coerce the visitor to disable their content blocking tool.
\section{Conclusion}
\label{sec:conclusion}

In this work we have presented the first accurate and fully automated system
for classifying whether applying a filter list rule to a website would break
the user-desirable features on that website. Past work has documented the
significant privacy, performance and security benefits of filter-list-based blocking, but this work only counts the ``benefits'' side of the ledger.
Absent a way of systematically predicting 
the ``costs'' of adding more privacy protections,
privacy research risks becoming detached from reality. Without a scalable
way of estimating compatibility risk, more blocking, more filtering, and more interventions will always look better.
If the usability costs are ignored, a broken system will always appear more private than
a functioning one.

We hope our work is a useful step towards finding practical, scalable ways of detecting
when privacy interventions break the systems they aim to improve. Our work
focuses on filter lists rules (because filter lists are among the most popular and
well-studied privacy interventions on the Web), but all proposed privacy interventions
would benefit from similar systems.

\bibliographystyle{IEEEtran}
\bibliography{paper}

\end{document}